\documentclass[twocolumn,aps,showpacs,prl]{revtex4}
\usepackage{graphicx}
\usepackage{epstopdf}
\usepackage{amsmath}
\usepackage{multirow}
\begin{document}

\title{Electronic and magnetic structures of
ternary iron selenides AFe$_2$Se$_2$ (A=K, Cs, or Tl)}

\author{Xun-Wang Yan$^{1,2}$}
\author{Miao Gao$^{1}$}
\author{Zhong-Yi Lu$^{1}$}\email{zlu@ruc.edu.cn}
\author{Tao Xiang$^{3,2}$}\email{txiang@aphy.iphy.ac.cn}

\date{\today}

\affiliation{$^{1}$Department of Physics, Renmin University of
China, Beijing 100872, China}

\affiliation{$^{2}$Institute of Theoretical Physics, Chinese Academy
of Sciences, Beijing 100190, China }

\affiliation{$^{3}$Institute of Physics, Chinese Academy of
Sciences, Beijing 100190, China }

\begin{abstract}

By the first-principles electronic structure calculations, we find
that the ground state of ternary iron selenides AFe$_2$Se$_2$ (A=K,
Cs, or Tl) is in a bi-collinear antiferromagnetic order, in which
the Fe local moments ($\sim2.8\mu_B$) align ferromagnetically along
a diagonal direction and antiferromagnetically along the other
diagonal direction on the Fe-Fe square lattice. This bi-collinear
antiferromagnetic order results from the interplay among the
nearest, the next nearest, and the next next nearest neighbor
superexchange interactions, mediated by Se $4p$-orbitals.

\end{abstract}


\maketitle

%\section{Introduction}

The discovery of high transition temperature $T_c$ superconductivity
in LaFeAsO by partial substitution of O with F atoms \cite{kamihara}
stimulates the intense studies on the iron pnictides. There have
been four types of iron-based compounds reported to show
superconductivity after doping or under high pressures, i.e.
1111-type $Re$FeAsO ($Re$ = rare earth) \cite{kamihara}, 122-type
$B$Fe$_2$As$_2$ ($B$=Ba, Sr, or Ca) \cite{rotter}, 111-type $A$FeAs
($A$ = alkali metal) \cite{wang}, and 11-type tetragonal
$\alpha$-FeSe(Te) \cite{hsu}. It was very recently reported that the
superconductivity was observed at about 30 K in a new FeSe-layer
compound K$_{0.8}$Fe$_2$Se$_2$ \cite{chen}, formed by intercalating
potassium (K) atoms between FeSe layers. Soon after, the
superconductivity was also found in the Cs-intercalated compound
Cs$_{0.8}$(FeSe$_{0.98}$)$_2$ \cite{Cs} and Tl-intercalated compound
TlFe$_x$Se$_2$\cite{fang}. Although these the compounds take the
ThCr$_{2}$Si$_{2}$ type structure with $P4/nmm$ symmetry
isostructural with 122-type $B$Fe$_2$As$_2$, they can be considered
as a new type of iron-based superconductors since they are
chalcogenides rather than pnictides.

For the parent compounds AFe$_2$Se$_2$ (A=K, Cs, or Tl) of these new
superconductors, the intercalated alkali metal or Tl atoms will
directly dope much more electrons into the FeSe layers than for the
iron pnictides. It is thus expected that the electronic and magnetic
structures of these new compounds are very likely different from the
ones of those iron pnictides. On the other hand, in order to
investigate the mechanism of superconductivity in these materials,
one needs to first understand the electronic and magnetic structures
of the parent compounds AFe$_2$Se$_2$.

In this paper, we report the theoretical result on the electronic
and magnetic structures of iron selenides AFe$_2$Se$_2$ (A=K, Cs, or
Tl) obtained from the first-principles electronic structure
calculations. We find that the compounds AFe$_2$Se$_2$ are
antiferromagnetic semimetals with a bi-collinear antiferromagnetic
order in the ground states, resulting from the strong nearest
($J_1$), next-nearest ($J_2$), and next-next-nearest ($J_3$)
neighbor superexchange interactions in these materials. A small
monoclinic lattice distortion due to spin-lattice coupling was
further found in these compounds, similar to $\alpha$-FeTe
\cite{ma,shi}. Here the bi-collinear antiferromagnetic (AFM) order
means that the Fe moments align ferromagnetically along a diagonal
direction and antiferromagnetically along the other diagonal
direction on the Fe-Fe square lattice. In other words, if the Fe-Fe
square lattice is divided into two square sublattices, the Fe
moments on each sublattice take their own collinear AFM order. The
bi-collinear antiferromagnetic order was first predicted for
$\alpha$-FeTe by our previous work \cite{ma} and confirmed by the
later neutron scattering experiment \cite{shi}.

\begin{figure}
\includegraphics[width=8.0cm]{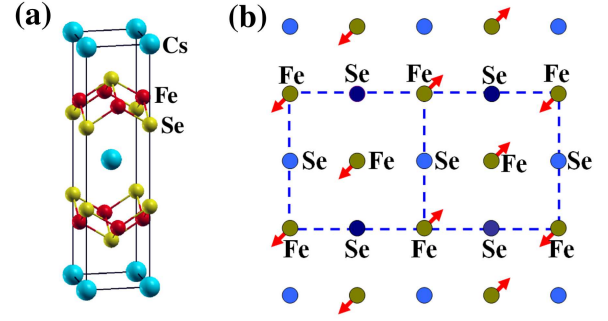}
\caption{(Color online) CsFe$_2$Se$_2$ with the ZrCuSiAs-type
structure: (a) a tetragonal unit cell containing two formula units;
(b) schematic top view of the FeSe layer. The large dashed square is
a $a\times 2 a$ unit cell. The Fe spins in the bicollinear
antiferromagnetic order are shown by the red arrows.} \label{figa}
\end{figure}

Although AFe$_2$Se$_2$ can be considered as a tetragonal crystal
with two formula units included in the corresponding unit cell as
shown in Fig.\ref{figa}(a), its primitive unit cell is constructed
by considering AFe$_2$Se$_2$ as a triclinic crystal, in which only
one formula unit cell is included. In the nonmagnetic calculations,
we adopted the primitive cell as a calculation cell. In the
calculation of the electronic and magnetic structures of the
ferromagnetic and the square antiferromagnetic Neel states, the
$a\times a$ FeSe cell is taken as the base cell. In the calculation
of the collinear and bi-collinear antiferromagnetic states, the unit
cells are doubled and the base cells are the
$\sqrt{2}a\times\sqrt{2}a$ and $a\times 2a$ (Fig. \ref{figa} (b))
FeSe cells, respectively.

In our calculations the plane wave basis method was used
\cite{pwscf}. We adopted the generalized gradient approximation
(GGA) of Perdew-Burke-Ernzerhof \cite{pbe} for the
exchange-correlation potentials. The ultrasoft pseudopotentials
\cite{vanderbilt} were used to model the electron-ion interactions.
After the full convergence test, the kinetic energy cut-off and the
charge density cut-off of the plane wave basis were chosen to be 800
eV and 6400 eV, respectively. The Gaussian broadening technique was
used and a mesh of $18\times 18\times 9$ k-points were sampled for
the Brillouin-zone integration. In the calculations, the
experimental tetragonal lattice parameters were adopted
\cite{chen,Cs}, and the internal atomic coordinates within the cell
were determined by the energy minimization. Actually, the lattice
parameters optimized by the energy minimization in our magnetic
calculations are found in excellent agreement with the experimental
ones.

\begin{figure}
\includegraphics[width=8.0cm]{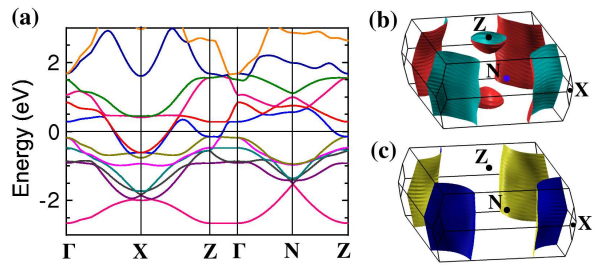}
\caption{(Color online) CsFe$_2$Se$_2$ in the nonmagnetic state: (a)
energy band structure; (b) the Fermi surface sheets due to the band
crossing the Fermi energy marked by the blue line in the left
figure; (c) the Fermi surface sheet due to the band crossing the
Fermi energy marked by the red line in the left figure.}\label{figb}
\end{figure}

In Fig. \ref{figb} we plot the electronic band structure and Fermi
surface of CsFe$_2$Se$_2$ in the nonmagnetic state, similar to the
ones just reported for KFe$_2$Se$_2$ \cite{shein}. As we see, there
are three Fermi surface sheets, among which the two cylinder-like
ones are located around the corners of the Brillouin zone, and
another one is of small pocket. The Fermi surface is all of
electron-type. The volume enclosed by these Fermi sheets give
$1.008$ electrons per formula cell. i.e. about $0.84 \times 10^{22}
/ cm^3$ . The density of states (DOS) at the Fermi energy is about
2.27 states per eV per formula unit. The corresponding electronic
specific heat coefficient and Pauli susceptibility are $\gamma =
5.35~mJ/(K^2\ast mol)$ and $\chi =0.92\times 10^{-9}~m^3 / mol$,
respectively. Our calculations for the nonmagnetic states exclude
any possible structural distortions. This suggests that any possible
structural distortion happening in CsFe$_2$Se$_2$ would be driven
through spin-phonon interactions. For KFe$_2$Se$_2$, the almost same
results are found by our calculations.

In order to explore the magnetic structure of CsFe$_2$Se$_2$, we
have calculated four different possible magnetic states with
ferromagnetic, square Neel AFM, collinear AFM, and bi-collinear AFM
orders, respectively. If the energy of the nonmagnetic state is set
to zero, we find that the energies of the ferromagnetic ($E_{FE}$),
square Neel AFM ($E_{AFM}$), collinear AFM ($E_{COL}$), and
bi-collinear AFM states ($E_{BI}$) are (-0.2122, -0.1651, -0.2677,
-0.2752) eV/Fe for CsFe$_2$Se$_2$ and (-0.2437, -0.1544, -0.2613,
-0.2884) eV/Fe for KFe$_2$Se$_2$, respectively. Thus the ground
states of both CsFe$_2$Se$_2$ and KFe$_2$Se$_2$ are in the
bi-collinear antiferromagnetic order, similar to the one of
$\alpha$-FeTe \cite{ma}. The magnetic moment around each Fe atom is
found to be about $2.4\sim 3.0~\mu_B$, varying weakly in the above
four magnetically ordered states, similar as in LaFeAsO and
BaFe$_2$As$_2$ \cite{ma1,ma2}. Since these local moments are
embedded in the environment of itinerant electrons, the moment of Fe
ions is thus fluctuating.

To quantify the magnetic interactions, we assume that the energy
differences between these magnetic orderings are predominantly
contributed from the exchange interactions between the Fe moments
with spin $\vec{S}$, which can be effectively modeled by the
following frustrated Heisenberg model with the nearest,
next-nearest, and next next nearest neighbor couplings $J_1$, $J_2$,
and $J_3$,
\begin{equation}\label{eq:Heisenberg}
H=J_1\sum_{\langle ij \rangle}\vec{S}_i\cdot\vec{S}_j +J_2\sum_{ \ll
ij \gg}\vec{S}_i\cdot\vec{S}_j +J_3\sum_{ \langle\langle\langle ij
\rangle\rangle\rangle}\vec{S}_i\cdot\vec{S}_j,
\end{equation}
whereas $\langle ij \rangle$, $\ll ij \gg$ and
$\langle\langle\langle ij \rangle\rangle\rangle$ denote the
summation over the nearest, next-nearest, and next-next-nearest
neighbors, respectively. This model may miss certain contributions
from itinerant electrons, however, we believe that it captures the
substantial physics on the magnetic structures. From the above
calculated energy data, we find that for CsFe$_2$Se$_2$
$J_1=-11.78~meV/S^2$, $J_2=19.75~meV/S^2$, and $J_3=11.75~meV/S^2$,
while for KFe$_2$Se$_2$ $J_1=-22.3 ~meV/S^2$, $J_2=15.6~meV/S^2$,
and $J_3=14.6~meV/S^2$ (The detailed calculation is referred to
Appendix of Ref. \onlinecite{ma1}). Notice that the ferromagnetic
states in both the compounds are lower in energy than the square
Neel AFM state, which results in the negative $J_1$ here.

It is known that when $J_3>J_2/2$ and $J_2>J_1/2$, the bi-collinear
AFM state is lower in energy than the collinear AFM state for a
frustrated $J_1$-$J_2$-$J_3$ Heisenberg model. This is in well
agreement with the derived $J_1$, $J_2$, and $J_3$ for both
CsFe$_2$Se$_2$ and KFe$_2$Se$_2$. Accordingly, by $J_1$-$J_2$-$J_3$
Heisenberg model we may understand the complex magnetic structures
displayed in both the compounds. On the other hand, the classical
study of $J_1$-$J_2$-$J_3$ Heisenberg model \cite{Ferrer,hu} shows
that there may be an incommensurate AFM spin order easily developed
in such $J_1$-$J_2$-$J_3$ system. However, it can be shown that a
slight structural distortion making $J_1$ different along different
direction will eliminate such an incommensurate spin order\cite{hu}.

It is expected that there would be a further lattice distortion
considering possible spin-phonon interactions. Similarly to
spin-Peierls distortion, the lattice constant slightly expands along
spin anti-parallel alignment to lower AFM energy and/or slightly
contracts along spin-parallel alignment to lower further
ferromagnetic energy. Indeed such small structural distortions are
found for both CsFe$_2$Se$_2$ and KFe$_2$Se$_2$ with an extra energy
gain of $\sim$4 meV/Fe. As a result, the crystal unit cell of the
compounds on FeSe layer deforms from a square to a rectangle
(rhombus). Such small lattice distortions affect weakly the
electronic band structures and the Fe moments.

Our calculations also show that for both CsFe$_2$Se$_2$ and
KFe$_2$Se$_2$ the Fe magnetic moments between the nearest neighbor
layers FeSe prefer the parallel alignment with a small energy gain
of about 2 meV/Fe in comparison with the anti-parallel alignment.
This is different from the iron pnictides \cite{ma,ma1,ma2}. It is
thus very likely that here the magnetic phase transition would
happen simultaneously with the structural transition. Overall the
magnetic order vector in these two compounds is thus
$(\frac{\pi}{a},0,0)$. The corresponding magnetic Bragg peaks are
$(1,0,0)$.

\begin{figure}
\includegraphics[width=9cm,height=4.5cm]{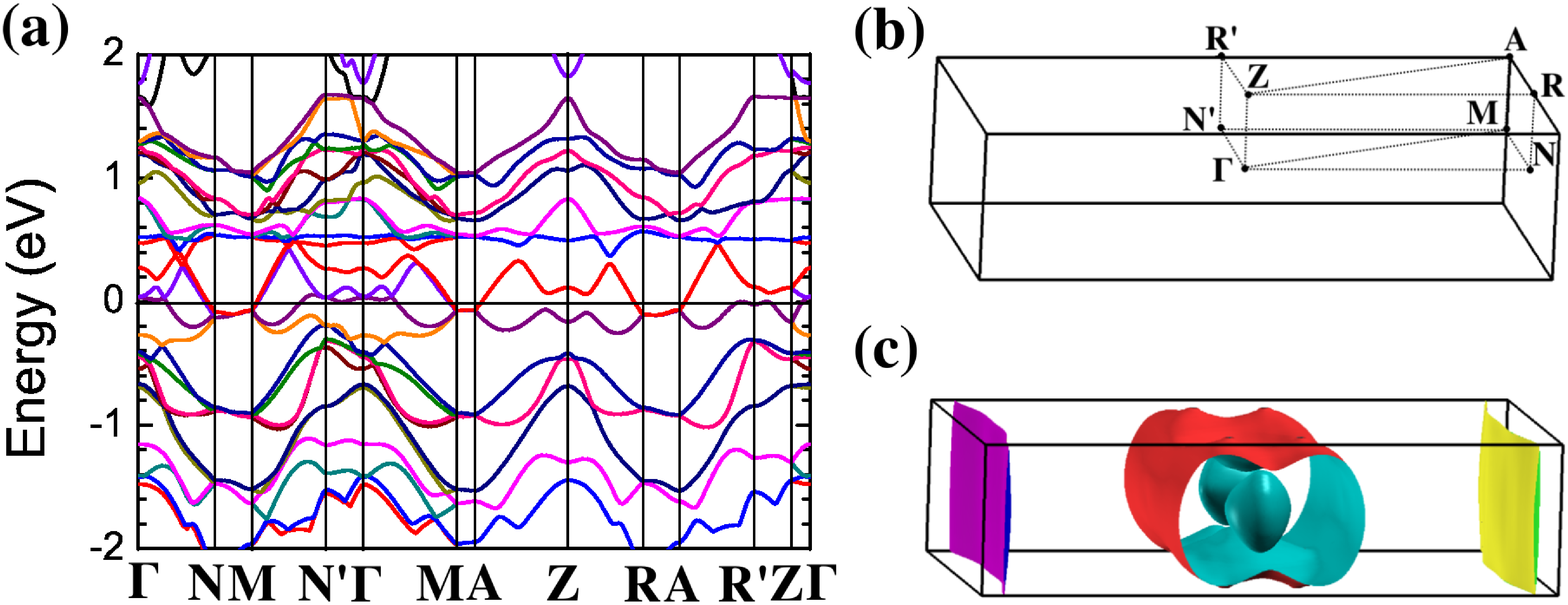}
\caption{(Color online) (a) Electronic band structure of
CsFe$_2$Se$_2$ in the bi-collinear-ordered antiferromagnetic state.
The Fermi energy is set to zero. (b) Brillouin zone. (c) Fermi
surface. Notice that $\Gamma$-N($\Gamma$-N$^{\prime}$) corresponds
the parallel(antiparallel)-aligned moment line. } \label{figc}
\end{figure}

Fig. \ref{figc} shows the electronic structure of CsFe$_2$Se$_2$ in
the bi-collinear AFM state, and the similar finding is for
KFe$_2$Se$_2$. There are three bands crossing the Fermi level which
form four sheets of the Fermi surface. The Fermi surface contains
two hole-type sheets parallel to the plane
$\Gamma$-Z-R$^{\prime}$-N$^{\prime}$, and two electron-type
cylinders parallel to the plane A-M-N-R. From the volumes enclosed
by these Fermi surface sheets, we find that the electron (hole)
carrier density is 0.052 electron/formula cell (0.061 hole/formula
cell), namely, $0.88\times 10^{21}/cm^3$ ($1.00\times
10^{21}/cm^3$). The density of states (DOS) at the Fermi level E$_F$
is 6.24 states per eV per formula cell. The corresponding electronic
specific heat coefficient $\gamma$ = 7.34 $mJ/(K^2*mol)$.

\begin{figure}
\includegraphics[width=6.5cm]{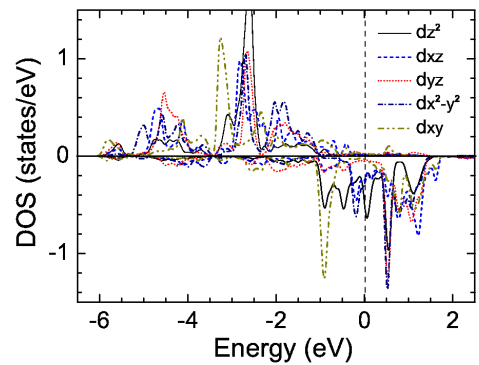}
\caption{(Color online) Calculated total and projected density of
states at the five Fe-$3d$ orbitals around one of the four Fe atoms
in the bi-collinear antiferromagnetic state unit cell of
CsFe$_2$Se$_2$. The Fermi energy is set to zero.}\label{figd}
\end{figure}

By projecting the density of states onto the five $3d$ orbitals of
Fe of CsFe$_2$Se$_2$ in the bi-collinear AFM state (see Fig.
\ref{figd}), we find that the five up-spin orbitals are almost
filled and the five down-spin orbitals are nearly uniformly filled
by half. This indicates that the crystal field splitting imposed by
Se atoms is very small. As the Hund rule coupling is strong, this
would lead to a large magnetic moment formed around each Fe atom, as
found in our calculations. The formation of Fe magnetic moments is
thus due to the Hund's rule coupling, which is a universal feature
found in all the iron-pnictides \cite{ma1,ma2}.

\begin{figure}
\includegraphics[width=6.5cm]{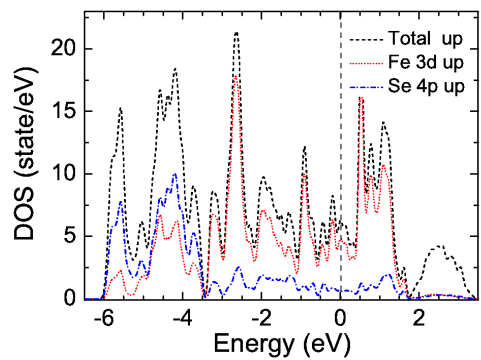}
\caption{(Color online) Total and orbital-resolved partial density
of states (spin-up part) per four formula cells of CsFe$_2$Se$_2$ in
the bi-collinear antiferromagnetic state. The Fermi energy is set to
zero.} \label{fige}
\end{figure}

Further inspection of the charge distribution in real space shows
that there is the covalence bond formed between the two nearest
neighbor Fe and Se atoms, which is responsible for the next nearest
neighbor Fe-Fe superexchange coupling $J_2$. This superexchange is
antiferromagnetic because the intermediated state associated with
the hopping bridged by Se atom is a spin singlet. Thus we find the
Se-bridged superexchange antiferromagnetic interaction between the
next nearest neighbor Fe-Fe atoms in CsFe$_2$Se$_2$ and
KFe$_2$Se$_2$, which is again similar to the ones in all the iron
pnictides \cite{ma1,ma2}. Nevertheless, there is a substantial
difference between CsFe$_2$Se$_2$ (KFe$_2$Se$_2$) and the iron
pnictides, regarding the next next nearest neighbor Fe-Fe exchange
interaction $J_3$ .

Fig. \ref{fige} shows that the band formed by Se $4p$ orbitals is
partially filled at the Fermi energy in CsFe$_2$Se$_2$ and
KFe$_2$Se$_2$. So there are itinerant $4p$ electrons at the Fermi
energy involved in mediating the exchange interactions in
CsFe$_2$Se$_2$ and KFe$_2$Se$_2$. This may explain why the next next
nearest neighbor superexhange coupling $J_3$ is large for
CsFe$_2$Se$_2$ and KFe$_2$Se$_2$ due to an RKKY-like mechanism. In
contrast, the band formed by As $4p$ orbitals is gapped at the Fermi
energy in the iron pnictides \cite{ma1,ma2}, correspondingly the
next next nearest neighbor superexchange coupling $J_3$ is nearly
zero.

A physical picture suggested in our study is thus that the Fe
moments in CsFe$_2$ and KFe$_2$Se$_2$ are also partially mediated by
partially delocalized Se $4p$-band and the origin of $J_3$ exchange
coupling may be well induced through an RKKY-like mechanism besides
the Se-bridged superexchange antiferromagnetic interaction $J_2$.
Meanwhile, the exchange interactions $J_2$ and $J_3$ are in strong
frustration competition, which is sensitively modulated by the Fe-Se
bonding and related local environment, as we see from $\alpha$-FeSe
($J_3$ $<$ $I_2$/2) \cite{ma} to CsFe$_2$Se$_2$ and KFe$_2$Se$_2$
here ($J_3$ $>$ $I_2$/2). By our calculation, it is clear that the
magnetic order should be along the diagonal direction of the Fe-Fe
square lattice for CsFe$_2$Se$_2$ and KFe$_2$Se$_2$ since $J_3$ is
larger than $J_2$/2. However, it is also very likely that defects,
vacancies, or other impurities would drive the bi-collinear
antiferromagnetic order into an incommensurate spin order along the
diagonal direction in CsFe$_2$Se$_2$ and KFe$_2$Se$_2$, similar to
the ones found in $\alpha$-FeTe \cite{bao}.

\begin{figure}
\includegraphics[width=6.5cm]{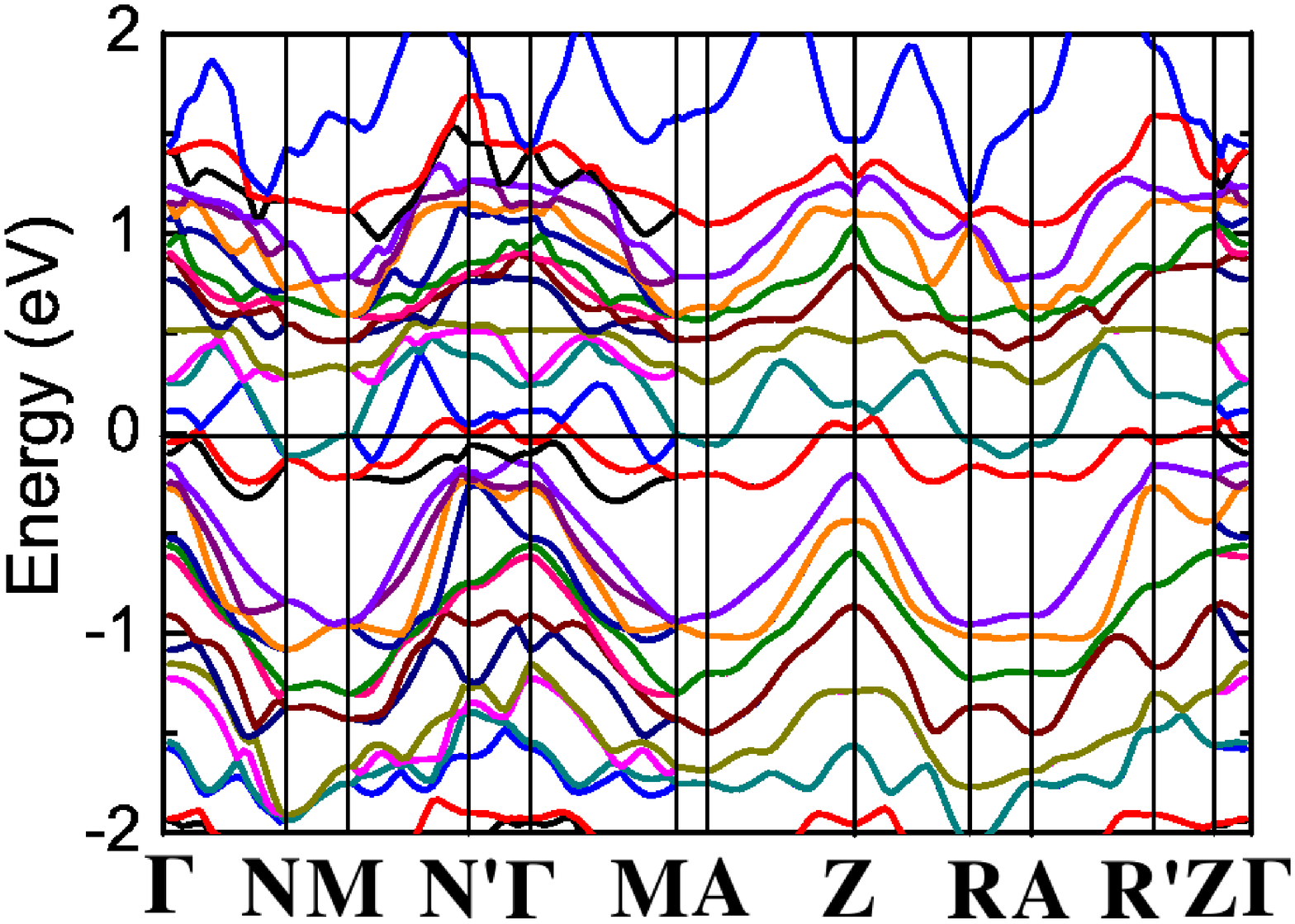}
\caption{(Color online) Electronic band structure of TlFe$_2$Se$_2$
in the bi-collinear-ordered antiferromagnetic state. The Brillouin
zone is shown in Fig. \ref{figc}(b). Notice that
$\Gamma$-N($\Gamma$-N$^{\prime}$) corresponds the
parallel(antiparallel)-aligned moment line. } \label{figf}
\end{figure}

Although Tl is not an alkali metal element, our calculations show
that the electronic and magnetic structures of TlFe$_2$Se$_2$ are
overall quite similar to the ones of CsFe$_2$Se$_2$ and
KFe$_2$Se$_2$. The electronic band structure and the Fermi surface
of TlFe$_2$Se$_2$ in the nonmagnetic state are found by our
calculations to be the same as the ones reported for TlFe$_2$Se$_2$
in Ref. \onlinecite{singh}. However, we find that the ground state
of TlFe$_2$Se$_2$ is in the bi-collinear AFM order with a magnetic
moment of $\sim2.7\mu_B$ around each Fe atom rather than the square
Neel AFM order reported in Ref. \onlinecite{singh}. The
corresponding electronic band structure is shown in Fig. \ref{figf}.

%\section{Conclusion}

In conclusion, we have presented the results of the electronic band
structure and magnetic properties of AFe$_2$Se$_2$ (A=K, Cs, or Tl)
based on the first-principles electronic structure calculations. our
studies show that the ground state of AFe$_2$Se$_2$ is a
quasi-2-dimensional bi-collinear antiferromagnetic semimetal with a
magnetic moment of $\sim2.8\mu_B$ around each Fe atom. This
bi-collinear antiferromagnetic state can be understood by the
Hesienberg model with $J_1$-$J_2$-$J_3$ superexchange interactions.

%\acknowledgements

This work is partially supported by National Natural Science
Foundation of China and by National Program for Basic Research of
MOST, China.


\begin{references}
\bibitem{kamihara}Y. Kamihara, T. Watanabe, M. Hirano, and H. Hosono,
J. Am. Chem. Soc. {\bf 130}, 3296 (2008).

\bibitem{rotter}M. Rotter, M. Tegel, and D. Johrendt, Phys. Rev. Lett. {\bf 101}, 107006
(2008).

\bibitem{wang}X.C.Wang, Q.Q. Liu, Y.X. Lv, W.B. Gao, L.X.Yang, R.C.Yu, F.Y.Li, and C.Q.
Jin, Solid State Commun. {\bf 148}, 538 (2008).

\bibitem{hsu} F.-C. Hsu, {\it et al.}, Proc. Natl. Acad. Sci.
{\bf 105}, 14262 (2008).

\bibitem{chen} Jiangang Guo, {\it et al.}, Phys. Rev. B {\bf 82}, 180520(R)
(2010).

\bibitem{Cs} A. Krzton-Maziopa, {\it et al.}, arXiv:1012.3637.
%Z. Shermadini, E. Pomjakushina, V. Pomjakushin,
%M. Bendele, A. Amato, R. Khasanov, H. Luetkens, K. Conder

\bibitem{fang} M. Fang, {\it et al.}, arXiv:1012.5236.
% Hangdong Wang1,2, Chiheng Dong1, Zujuan Li1, Chunmu Feng1, Jian Chen1, H.Q. Yuan1

\bibitem{ma} F. Ma {\it et al.}, Phys. Rev. Lett. {\bf 102}, 177003 (2009).

\bibitem{shi} S.L. Li, {\it et al.}, Phys. Rev. B {\bf 79}, 054503 (2009).


%\bibitem{ma} F. Ma and Z.-Y. Lu, Phys. Rev. B {\bf 78}, 033111 (2008).


\bibitem{pwscf}P. Giannozzi, {\it et al.}, http://www.quantum-espresso.org.

\bibitem{pbe}J. P. Perdew, K. Burke, and M. Erznerhof,
Phys. Rev. Lett. {\bf 77}, 3865 (1996).

\bibitem{vanderbilt}D. Vanderbilt, Phys. Rev. B {\bf 41}, 7892 (1990).

\bibitem{shein}I.R. Shein and A.L. Ivanovskii, arXiv:1012.5164.

\bibitem{ma1}F. Ma, Z.Y. Lu, and T. Xiang, Phys. Rev. B {\bf 78}, 224517
(2008).

\bibitem{ma2} F. Ma, Z.Y. Lu, and T. Xiang, Front. Phys. China, {\bf 5(2)}, 150
(2010).

\bibitem{Ferrer}J. Ferrer, Phys. Rev. B  {\bf 47}, 8769 (1993).

\bibitem{hu} C. Fang, B. A. Bernevig, and J. Hu, Europhys. Lett. {\bf 86}, 67005
(2009).

\bibitem{bao} W. Bao {\it et al.}, Phys. Rev. Lett. {\bf 102}, 247001 (2009).

\bibitem{singh} L. Zhang and D.J. Singh, Phys. Rev. B {\bf 79}, 094528
(2009).

\end{references}
\end{document}